\documentclass[seceq]{ptptex}

\usepackage{graphicx}

\newcommand{\beq}{\begin{equation}}
\newcommand{\eeq}{\end{equation}}
\newcommand{\beqa}{\begin{eqnarray}}
\newcommand{\eeqa}{\end{eqnarray}}
\newcommand{\bmat}{\begin{displaymath}}
\newcommand{\emat}{\end{displaymath}}

\newcommand{\eq}[1]{Eq.~(\ref{#1})}

\title{
From the Frenkel-Kontorova model to Josephson junction arrays\\

}
\subtitle{
the Aubry's transition as a jamming-glass transition} 

\author{
Hajime \textsc{Yoshino}$^1$
Tomoaki  \textsc{Nogawa}$^2$ and
Bongsoo \textsc{Kim}$^3$%
}

\inst{
$^1$Department of Earth and Space Science, Faculty of Science,
 Osaka University, Toyonaka 560-0043, Japan\\
$^2$Department of Applied Physics. School of Engineering, 
The University of Tokyo, 7-3-1 Hongo, Bunkyo-ku, Tokyo 113-8656, Japan\\
$^3$Department of Physics, Changwon National University, 
Changwon 641-773, Korea.
}

\abst{
The Frenkel Kontorova (FK) model is known to exhibit the so called
Aubry's transition which is a jamming or frictional transition at zero
temperature. 
Recently we found similar transition at zero and finite temperatures
in a super-conducting Josephson junction array (JJA) on a square lattice
under external magnetic field.
In the present paper we discuss how these problems are related. 
}

\begin{document}

\maketitle

\section{Introduction}

Understanding of non-crystalline solids 
such as glasses and granular systems is an important
problem in condensed matter physics \cite{Jamming-Book,kurchan-levine}. 
A useful concept is that {\it frustration} of geometrical, energetic
or kinetic origins is indispensable to avoid crystallization and allow
realization of amorphous solids \cite{Sadoc,Tarjus-review}.
In the present paper we discuss a jamming in a strongly frustrated
Josephson junction array (JJA) under external magnetic field
\cite{Yoshino-Nogawa-Kim-1,Yoshino-Nogawa-Kim-2,Yoshino-Nogawa-Kim-3}.  
It is a very interesting
system which provides an exceptional opportunity to study both athermal
(jamming) and thermal (glass) transitions in exactly the same settings. The
question raised by the Chicago group - whether athermal and thermal
jamming or glass transitions can be understood in a unified way
\cite{Jamming-Book,Liu-Nagel,Nagel-group} - can be asked explicitly in
this system. 

In the present paper we discuss the possibility that athermal and thermal jamming transition in the present system can be understood as a generalization
of the Aubry's transition \cite{Aubry-Daeron,Peyrard-Aubry} found in a
family of one-dimensional models of {\it frictions}, most importantly
the  Frenkel-Kontorova (FK) model which exhibits very rich phenomenology
in spite of its simplicity \cite{FK-review}.

The organization of the paper is as follows.  In the next section, we
discuss the sequence of connections between the FK model \cite{FK-review},
Matsukawa-Fukuyama (MF) model \cite{Matsukawa-Fukuyama} and the
frustrated Josephson junction array under magnetic field
\cite{Tinkam,Mooij-group,Martinoli-Leemann} step by step. In sec. \ref{sec-aubry} we review the Aubry's transition
\cite{Aubry-Daeron,Peyrard-Aubry} in the FK model. There we focus on the
properties of the so called hull function which is a powerful
theoretical tool to analyze the Aubry's transition. Then we sketch our
recent attempt to generalize it for the case of frustrated JJA \cite{Yoshino-Nogawa-Kim-3}. In sec.~\ref{sec-shear} we point out that
'shear' can be exerted on JJA via external electric current \cite{Yoshino}.
We discuss how tribology (sliding friction) \cite{Persson}, non-linear rheology
(soft-matters, granular matters, e.t.c.)
\cite{rheologybook,Otsuki-Sasa,Hatano-Otsuki-Sasa,Olsson-Teitel,Hatano,Otsuki-Hayakawa}
and non-linear transport (JJA,
superconductors, e.t.c) \cite{Tinkam,FFH} are related to
each other emphasizing remarkable similarity of their scaling
features around critical points including the J (Jamming)-point.
Finally we discuss the ``Jamming phase diagram'' of the JJA, which is
analogous to the one proposed for soft-matters
\cite{Liu-Nagel,Nagel-group},  suggested by our analysis of non-linear transport properties
at zero temperature \cite{Yoshino-Nogawa-Kim-1} 
and Monte Carlo simulations at finite temperatures \cite{Yoshino-Nogawa-Kim-2}.
In sec. \ref{sec-conclusions} we summarize this paper and
discuss some future outlooks.

\section{Link between the friction models and the Josephson junction
 arrays}
\label{sec-link}

\begin{figure}[h]
\begin{center}
\includegraphics[width=0.5\textwidth]{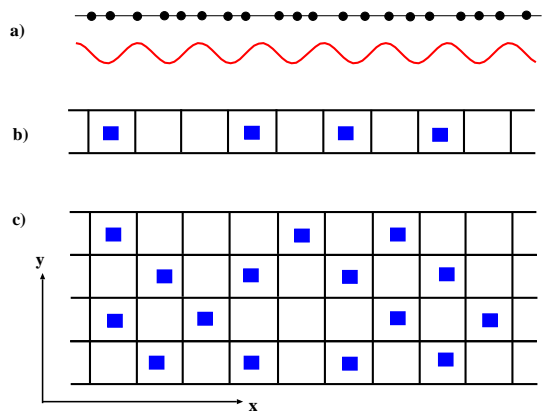}
\end{center}
\caption{(Color online) Schematic pictures of the friction models and the Josephson
 junction array. a) the Frenkel-Kontorova (FK) model b)
 Matsukawa-Fukuyama's 2-chain model on the ladder lattice 
and c) the Josephson junction array (JJA) on a square lattice under
 external magnetic field.
The squares in the plaquette represent positions of the vortexes
induced either by mismatching between the two incommensurate surfaces b) or
external magnetic field c).
} 
\label{fig-fk-jja}
\end{figure}

\subsection{Frenkel-Kontorova model - starting point}

\subsubsection{Frustration due to mismatching}

The original Frenkel-Kontorova model \cite{FK-review} is a one-dimensional elastic chain of
particles put on a periodic substrate (see Fig.~\ref{fig-fk-jja} a)). The Hamiltonian is given by,
\beq
H=\sum_{n=1}^{L} \left\{\frac{k}{2}(u_{n+1}-u_{n}-l)^2
- \lambda \cos \left(\frac{\pi u_{n}}{a}\right) \right\}.
\eeq
Here $u_{n}$ denotes the position of the $n$-th particle. The particles are
connected to each other by Hookian springs of strength $k$ as described by the 1st term
in the Hamiltonian where $l$ is the natural spacing between the
particles. We impose a boundary condition such that the length 
of the whole system is fixed $(u_{L+1}-u_{1})/l=L$.\footnote{ The FK
model with fixed volume (length) and that under fixed pressure (external
force) behave completely differently \cite{FK-review,Peyrard-Aubry}. 
The former is relevant in the context of friction (jamming) and
	frustration is in some sense stronger than the latter. In the
	latter case the response of the system with respect to the
	increments of the external force exhibits devil's stair case
	singularities. Note also that the Hookian spring force, which arises 
due to the harmonic potential, does not endure the natural spacing $l$
	by itself.}
The 2nd term 
describes the periodic potential due to the substrate whose period is $a$.

There are two important parameters: 1) the strength of the potential $\lambda$ 
and 2) winding number $f$,
\beq
f=\frac{l}{2a}.
\eeq
Both are crucial for the jamming-unjamming transition (Aubry's
transition) in the FK model \cite{Aubry-Daeron,Peyrard-Aubry}.
Later we will find equivalent two parameters in the frustrated Josephson
junction array under external magnetic field.

The sinusoidal potential allows the elastic chain to make
phase slips with respect to the substrate. Thus this simple system
allows both elastic and plastic deformations.
The elastic term prefers to keep the natural spacing $l$ while the
substrate potential prefers $2a$. 
In the context of the friction between two different materials brought
in contact with each other \cite{Persson}, 
it is natural to suppose that the two surfaces are {\it incommensurate} 
with respect to each other, namely $f$ is an {\it irrational} number -
a number which cannot be represented as a ratio of some two integers.
As the result the system becomes {\it frustrated}
as soon as $\lambda$ becomes finite. 
Finding the ground state of the system, which is some periodic (possibly
of very long periodicity) crystalline
structure for rational $f$, becomes a highly non-trivial problem \cite{ground-states-isotropic}.
In the present paper we always assume $f$ is irrational.
\footnote{For technical reasons we wish to use periodic boundary conditions which
cannot be compatible with {\it irrational} $f$. Thus in practice 
we use rational numbers which approximates a target irrational number.
For instance we can take a series of integers $p_{n}$ with
$n=1,2,\ldots$ from the Fibonacci series and construct a series of 
rational numbers $p_{n-1}/p_{n}$ which converges to
$f=(3-\sqrt{5})/20.38196601...$ in the limit $n \to \infty$.
We consider systems with linear size $L=p_{n}$ so that we arrive at the
target irrational number in the thermodynamic limit $L \to \infty$.
Note that it is easy to construct similar Fibonacci-like series for any
irrational numbers which are solutions of some quadratic equations
\cite{Yoshino-Nogawa-Kim-1}.}

The system exhibits a jamming or frictional transition - called as
Aubry's transition \cite{Aubry-Daeron,Peyrard-Aubry} which we review in section \ref{sec-aubry}.
For weak enough coupling $\lambda < \lambda_{c}$
the elastic chain is only mildly deformed and it can slide over the
substrate smoothly without energy dissipation - friction-less.
For stronger coupling $\lambda > \lambda_{c}$, the elastic chain becomes
pinned by the substrate and friction emerges. We will find later that
the parameter $\lambda$, which plays a key role in the FK model,
 is equivalent to strength of anisotropy of the Josephson coupling in the Josephson junction array.

\subsubsection{Phase representation}

It is convenient to introduce a dimension-less ``phase'' variable $\theta_{n}$
defined by $u_{n}=\frac{a}{\pi}[\theta_{n}+2\pi (f-1)n]$ by which 
the (dimension-less) Hamiltonian can be rewritten as,
\beq
H=\sum_{n=1}^{L} \left\{\frac{1}{2}(\theta_{n+1}-\theta_{n}-2\pi)^2
- \lambda\cos (\theta_{n}+2\pi f n)\right\}
\label{eq-FK-phase}
\eeq
where $\lambda$ is also made dimension-less by an appropriate rescaling. The
boundary condition is such that $(\theta_{L+1}-\theta_{1})/2\pi=L$ is
fixed.

\subsection{Matsukawa-Fukuyama model - a crucial intermediate step}

\subsubsection{Phase model on ladder-lattice}

Matsukawa and Fukuyama considered a two-chain model in the context of
friction \cite{Matsukawa-Fukuyama}. Their idea is to allow the
``substrate'' in the FK model to deform elastically as well, which is
certainly more realistic than the FK model in the context of tribology \cite{Persson}. Following their idea, let us
modify the FK model \eq{eq-FK-phase} and develop a phase model defined
on a two-leg ladder lattice shown in Fig.\ref{fig-fk-jja} b). The hamiltonian is given by,
\beq
H= \sum_{\vec{e}_{ij}=\vec{e}_{x}} \frac{1}{2}(\theta_{i}-\theta_{j}-2\pi)^{2}
 - \lambda \sum_{\vec{e}_{ij}=\vec{e}_{y}}
 \cos(\theta_{i}-\theta_{j}-2\pi f n_{i})
\label{eq-2-chain-original}
\eeq
To simplify notations we relabeled the sites as $i=1,2,\ldots,N$ 
whose position in the real space is given by $(n_{i},m_{i})$.  In the two-chain model
the index for the column takes values $n=1,2,\ldots,L$ while that for the row (or layer) takes 
just two values $m=1,2$.
The sums are took over nearest neighbour pairs connected by
displacement vector $\vec{e}_{ij}=(n_{i}-n_{j},m_{i}-m_{j})$
which is either equal to $(1,0)$ or $(0,1)$.

\subsubsection{Gauge invariance}
\label{subsubsec-gauge}

An important property of the system is {\it gauge invariance} which we
explain below. Let us rewrite the hamiltonian \eq{eq-2-chain-original} as,
\beq
H= \sum_{\vec{e}_{ij}=(1,0)} \frac{1}{2}\psi^{2}_{ij}
 - \lambda \sum_{\vec{e}_{ij}=(0,1)} \cos(\psi_{ij})
\label{eq-2-chain}
\eeq
with the phase difference
\beq
\psi_{ij} \equiv \theta_{i}-\theta_{j}-A_{ij}.
\label{eq-def-psi}
\eeq
Here $A_{ij}$ is an anti-symmetric matrix $A_{ij}=-A_{ji}$ which satisfy,
\beq
\sum_{\rm plaquette} A_{i,j}=2\pi f.
\label{eq-vector-potential}
\eeq
The sum $\sum_{\rm plaquette}$ is a directed sum over ``bonds''
along each ``plaquette'' in the anti-clockwise manner.  

It is easy to see that the original representation
\eq{eq-2-chain-original} respect the condition
\eq{eq-vector-potential}. 
The crucial point is that the phase differences 
$\psi_{ij}$ are invariant under gauge transformations;
\begin{eqnarray}
&& \theta_{i} \to\theta_{i}+\delta \theta_{i}\\
&& A_{ij} \to A_{ij}+\delta \theta_{i} -\delta \theta_{j}
\end{eqnarray}
Thus the hamiltonian \eq{eq-2-chain-original} is gauge-invariant.
The condition \eq{eq-vector-potential} itself is also gauge-invariant.

In addition to the gauge invariance, the hamiltonian \eq{eq-2-chain}
is invariant under $A \to -A$ with $\theta_{n} \to -\theta_{n}$. 
Furthermore also $f \to 1+f$ does not change the problem. 
So we only need to consider $0 < f \leq 1/2$ in the following.

\subsection{Frustrated Josephson-junction array (JJA) under magnetic field}

\subsubsection{Frustration due to external magnetic field}

The final step is just to 1) increase the number of legs of the ladder to build a 2-dimensional square lattice and 2) replace the intra-layer elastic couplings by sinusoidal couplings ((see Fig.~\ref{fig-fk-jja} c)). Then we obtain the Josephson junction array on a square lattice under external magnetic field applied perpendicularly to the array \cite{Tinkam},
\beq
H= -\sum_{\vec{e}_{ij}=(1,0)} \cos(\psi_{ij})
 - \lambda \sum_{\vec{e}_{ij}=(0,1} \cos(\psi_{ij})
\label{eq-JJA-hamiltonian}
\eeq

Here $\theta_{i}$ is identified as the phase of the superconducting order parameter of the $i$-th site (superconducting island). The sinusoidal couplings represent Josephson coupling between the superconducting islands. Now the potential $A$ is identified as the vector potential due to external magnetic field $B_{z}$ applied along the $+z$ direction. The parameter $f$ which appears in \eq{eq-vector-potential} is the number density of quantized flux lines $f=a^{2}B_{z}/\phi_{0}$ where $a^{2}$,$B_{z}$ and $\phi_{0}$ are the are of the plaquette, strength of the magnetic and flux quantum. 

Let us emphasize that the two important parameters in the FK model, namely  1) the parameter $\lambda$ and 2) the winding number $f$ are inherited  down to the the JJA. To conclude we finally arrived a Josephson junction array on a square lattice with {\it anisotropic} coupling - with anisotropy $\lambda$ - under external magnetic field $B_{z}=(\phi_{0}/a^{2})f$ - with irrational number density $f$ of fluxes per plaquette. In short, let us call such a system as {\it irrationally frustrated  anisotropic} JJA.

Quite interestingly it is actually possible to construct {\it anisotropic} JUL in laboratory. The strength of the Josephson coupling depends, for instance, on the thickness of the junctions. Saito and Osada \cite{Saito-Osada} created anisotropic JJA with various $\lambda$ by controlling the thickness of the junctions in the lithography process.

It may sound rather strange to consider the {\it anisotropy} seriously
since it usual plays only minor roles. Not surprisingly, previous
studies of irrationally frustrated JJA considered only isotropic systems
$\lambda=1$.\footnote{We note however that Denniston and Tang
\cite{Denniston-Tang-ladder} studied the frustrated JJA on the
ladder-lattice (with $m=1,2$) (See Fig.~\ref{fig-fk-jja} b)) and
consider variation of the inter-leg coupling $\lambda$.  Their system 
is almost the same as the 2-chain model by Matsukawa and Fukuyama but the elastic intra-chain coupling in \eq{eq-2-chain} is replaced by a
sinusoidal coupling. They found the Aubry's transition also exist in the
frustrated JJA on the ladder.} 
 As we discuss later, it turned out in our recent studied that $\lambda$ is actually {\it relevant for irrational} $f$ \cite{Yoshino-Nogawa-Kim-1,Yoshino-Nogawa-Kim-2,Yoshino-Nogawa-Kim-3}. Quite remarkably the isotropic point $\lambda=1$ turned out to be a critical point at zero temperature corresponding to $\lambda_{c}$ of the FK model where a jamming transition analogous to the Aubry's transition takes place. By symmetry it is obvious that we only need to consider the case $\lambda \geq 1$.

\subsubsection{Vortex - analogue of dislocation}
\label{subsubsec-vortex}

We mentioned above that the parameter $f$ can be regarded as number
density of quantized flux lines per plaquette. As we explain below, this is
because the vector potential $A$ due to the magnetic field induces
vortexes of the phases $\theta_{i}$. The point is that vortexes are quantized objects like dislocations in crystals.

Here it is convenient to define ``charges'' of the vortexes as,
\beq
q_{i}=\frac{1}{2\pi}\sum_{\rm plaquette} s(\psi_{ij})=p_{i}-f \qquad
p_{i}=\ldots,-2,-1,0,1,2,\ldots
\eeq
where $s(x)$ is a saw-tooth like periodic function with period $2\pi$
and $s(x)=x$ in the range $-\pi < x \leq \pi$. By definition, the charge
$q$ takes only discrete values of the form $p_{i}-f$ with some 
integer $p_{i}$ and offset $-f$ as shown above.  
Physically the integer $p_{i}$ represents the number of quantized
fluxes (each carrying a flux quantum $\phi_{0}$) threading the $i$-th plaquette. 
Note also that the charge $q$ is gauge invariant.

The usefulness of the charge becomes manifested in the so called
coulomb-gas mapping (see Chap. 9 of \citen{Chaikin-Lubensky}) in which continuous, elastic deformations (``spin-wave'') 
are integrated out to find effective hamiltonian ${\cal H}$ of the vortexes.
The resultant system is essentially equivalent to a lattice-gas of electrostatic
charges interacting with each other by the repulsive coulomb interactions,
\beq
{\cal H}= \sum_{i \neq j} q_{i}G(\vec{r}_{ij})q_{j} +G({\vec 0})\sum_{i} q^{2}_{i}
\label{eq-coulomb-gass}
\eeq
with ${\vec r}_{ij}=(n_{i}-n_{j},m_{i}-m_{j})$. 

The interaction potential $G({\vec r})$ is the (static) Green's function
of elastic deformations (spin-wave). In 2-dimension, it scales as
$G({\vec r}) \propto \log(|{\vec r}|)$ for $r \gg 1$. 
Note that the anisotropy $\lambda$ in \eq{eq-JJA-hamiltonian} is simply reflected in
anisotropy in $G({\vec r})$ such that with it is stronger into
$y$-direction ${\vec r}|| (0,1)$ compared to $x$-direction ${\vec r}||
(1,0)$ by factor $\lambda (\geq 1)$.

The value $G({\vec 0})(>0)$ can be interpreted as the {\it core energy}
of the vortexes. 
Because of the core energy, states with higher values of the vortex
charges generally have larger energies and can be neglected at low
temperatures. Since we only need to
consider $0 < f \leq 1/2$ as noted in sec \ref{subsubsec-gauge},  
it is sufficient to consider two values of the charges $q=-f,1-f$.
In addition we assume the charge neutrality $\sum_{i}q_{i}=0$ holds,
which can be enforced by applying the periodic boundary conditions.
As the result we find that a fraction
$f$ of the plaquettes carries a vortex $p=1$ (or $q=1-f$) 
and the other fraction $1-f$ carries no vortex $p=0$ ( or $q=-f$).
In Fig.~\ref{fig-fk-jja}, the boxes in the plaquette represent the
vortexes ($p=1$).

\subsubsection{Vortex patterns in equilibrium - vortex liquid, crystal and glass}

Let us sketch briefly possible patterns of vortexes in equilibrium states
at low temperatures. For clarity we discuss three cases 1) $f=0$ 2) $f$
is rational and 3) $f$ is irrational.

If $f=0$, the ground state of the system is trivial: the phase becomes
uniformly ordered $\theta_{i}={\rm constant}$ for all sites $i$.  
In such a ground state the vortex is absent everywhere $p_{i}=0$ ($q_{i}=0$).
It can be regarded as a crystalline state (or ferromagnetic state). At
finite temperatures, pairs of vortex ($p=1$) and anti-vortex ($p=-1$)
will be created leading to melting of the crystalline state by
proliferation of the vortexes (and anti-vortexes) at some
critical temperature $T_{\rm c}$. In 2-dimension, it takes place 
in a special way named as Kosterlitz-Thouless transition \cite{KT}.

If $f$ is {\it rational}, i.~e. $f=p/q$
with some integers $p$ and $q$, the system will have a period {\it
vortex lattice} \cite{ground-states-isotropic}, which is analogous to
periodically ordered structure of dislocations in the so called Frank-Kasper phase
\cite{Frank-Kasper}. For example with $f=1/2$, the charges exhibit a checkerboard like order in which 
the sign of the charges alternates along $x$ and $y$-axis
as  $q=1/2,-1/2,1/2,-1/2,\ldots$ (or $p=1,0,1,0,\ldots$). We also note
that the 'half-vortexes' which appears in the case of $f=1/2$ is
identical to the so called {\it chirality} in frustrated magnets
\cite{Villain,TJ83-FFXY}.

In bulk superconductors formation of the vortex lattice is well
known. The latter is a triangular lattice called as Abrikosov lattice
\cite{Tinkam}.  On the other hand, the vortex lattices in JJA are formed on top of the
underlying square lattice so that it is a
{\it super}-lattice. Thus the vortex lattices in JJA are usually {\it pinned}
by the underlying lattice of the JJA while those in the bulk
{\it pure} superconductors are free to move around unless some pinning
centers are present \cite{FFH,collective-pinning-vortex-glass}.

Starting from the FK model we are naturally lead to consider {\it irrational} $f$.
Apparently the system cannot develop simple periodic
vortex lattices with irrational $f$ so that 
finding the ground state becomes a highly
non-trivial problem. Indeed JJA with irrational $f$ - {\it irrationally
frustrated JJA} - has been 
regarded as a system which possibly exhibit a glassy phase since a
seminal work by T. Halsey \cite{Halsey}.  This is a quite intriguing
possibility since it means emergence of a glassy phase with frustration
but {\it without quenched disorder}  - at variance with the conventional
spin-glasses and vortex-glasses (superconductors with random pinning
centers) which involve quenched disorder
\cite{FFH,collective-pinning-vortex-glass}. Disorder may be somehow self-generated in this system. Indeed equilibrium relaxations of the irrationally frustrated JJA were similar to the primary relaxation observed in typical fragile supercooled liquids \cite{JJA-relaxation}.

\section{Low lying states and Aubry's transition}
\label{sec-aubry}

\subsection{Hull function of the FK model}

\begin{figure}[b]
\begin{center}
\includegraphics[width=0.5\textwidth]{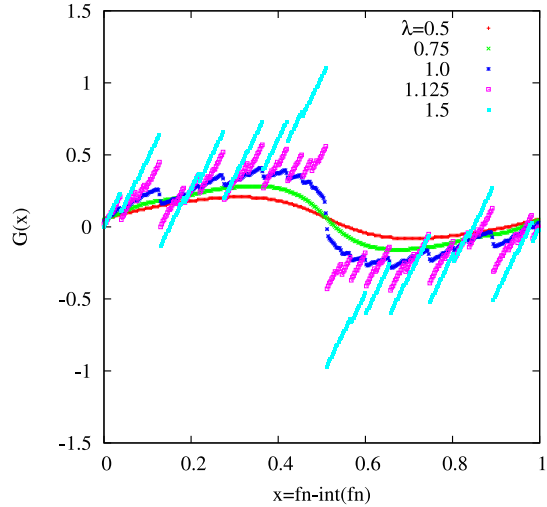}
\end{center}
\caption{(Color online) Hull function $G(x)$ of the FK model.
Here $\theta_{n}-2\pi n$ of the ground states
are plotted against the ``folded
 coordinate'' $[n]=fn-{\rm int}(fn)$ to elucidate the hull function.
$\lambda_{c}=0.9716..$ \cite{Peyrard-Aubry}.
} 
\label{fig-hull-FK}
\end{figure}

Now let us turn to review the Aubry's transition found in the FK model\cite{FK-review,Aubry-Daeron,Peyrard-Aubry,Zhirov}
and related friction models including the MF model \cite{Matsukawa-Fukuyama,kawaguchi-matsukawa}.  A remarkable feature of the FK model is that 
mathematically rigorous analysis of the low lying states is possible based on the fact that configuration of the energy minima (and
maxima) of the system satisfies a recursion relation which is identical
to the so called standard map well known in dynamical systems.

It is known rigorously that the ground state of the FK model can be expressed as\cite{Aubry-Daeron,Peyrard-Aubry},
\beq
\theta_{n}=2\pi n + G( f n + \alpha)
\label{eq-hull-FK}
\eeq
where $G$ is a periodic function with periodicity $1$, i.e. $G(x+1)=G(x)$
for any $x$. The function $G(x)$ is called  'hull function' and
describes distortion of the configuration of the elastic chain due to
the substrate potential. It is important to note that the entire region
$0 < x \leq 1$ becomes equally populated in the thermodynamic limit $L \to \infty$
for {\it irrational} $f$.

Quite remarkably the phase $\alpha$ is arbitrary, meaning that there is
a manifold of ground states which have exactly the same energy. 
Moreover it is known rigorously that there is a ``phase
transition'' for irrational $f$, called  'transition by breaking of
analycity' (or Aubry's transition),  at a critical strength of coupling
$\lambda_{c}(f)$. The basic feature of the transition is summarized in the table \ref{table-aubry}.
In Fig.~\ref{fig-hull-FK} we show the hull function of the FK model
constructed from numerically generated ground states at various
$\lambda$ (see Ref \cite{Peyrard-Aubry} for the method).

\begin{table}[t]
\begin{tabular}{c||c|c}
& $G(x)$ & (meta)stable states\\ \hline
$\lambda < \lambda_{c}(f)$ & analytic & only the ground state 
\\
$\lambda > \lambda_{c}(f)$ & non-analytic,
 with infinitely many discontinuities & infinitely many metastable states\cite{Zhirov}
\end{tabular}
\label{table-aubry}
\caption{Changes of the low lying states by the Aubry's transition}
\end{table}

From a physical point of view, a significant consequence of the Aubry's
transition is the ``frictional transition'' between the sliding phase and jamming (pinned) phase \cite{Peyrard-Aubry}. Let us sketch the essence of the
reasoning in the following.

For $\lambda < \lambda_{c}$, starting from a ground state, one can find
a continuum of states with exactly the same energy by varying
$\alpha$. The point is that they are all related to each other by some
continuous displacements of the particles in the real space. 
This comes 
from the fact that $G(x)$ is analytic for $\lambda < \lambda_{c}$. Thus
no external force is needed to {\it slide} the whole system - {\it
friction-less} or {\it sliding}. 

Existence of the sliding becomes trivial if the elastic
chain itself is replaced by a rigid body. In such an extreme case of
friction between two incommensurate rigid bodies, the forces between them
oscillates in the space with an incommensurate 
period so that the net force becomes cancelled out. 
The non-trivial point is
that similar cancellation of the forces still happens even if the chain
is allowed to deform elastically as long as the coupling 
$\lambda$ is sufficiently small.

For $\lambda > \lambda_{c}$, discontinuous points appear in the hull function $G(x)$. It means that variation of $\alpha$ require discontinuous movements of the particles in the real space.
Thus the ground states are no-more connected to each other by sliding:
the system prepared in the ground state has to go over some higher
energy states (thus energy barriers) to reach another ground state. Thus the system is
jammed. Now some finite strength of external force greater than a
certain frictional force (yield stress) $f_{\rm yield}(\lambda) \propto
(\lambda-\lambda_{c})^{\beta}$ must be applied to the system to let it
move (de-pinning) \cite{Peyrard-Aubry}. 

Firov et al \cite{Zhirov,FK-review} has been able to find a hierarchy of
exponentially large number of low lying states on top of the ground
state in the jammed phase $\lambda > \lambda_{\rm c}$. 
This is a very interesting observation from the view point
of the physics of glasses. However, unfortunately the FK model is an
one-dimensional system so that the Aubry's transition disappears at finite temperatures.

The frictional transition and emergence of discontinuity in the hull function has also been found in the Matsukawa-Fukuyama's 2-chain model \cite{kawaguchi-matsukawa}.
Now it is very natural to expect that these features will be inherited
down to our JJA on the square lattice under magnetic field. The main
message that we find here is that we should vary the anisotropy
$\lambda$ and see what happens in the low lying states.

\subsection{Low lying states of the anisotropic JJA}

\begin{figure}[t]
\begin{center}
\includegraphics[width=0.5\textwidth]{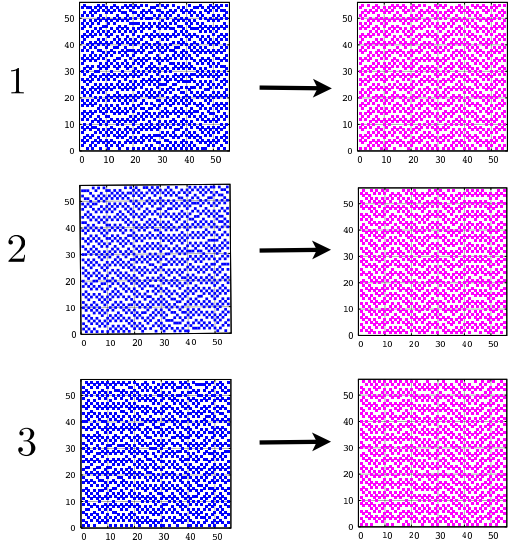}
\end{center}
\caption{(Color online) Undulated vortexes in
the anisotropic irrationally frustrated JJA. a) Examples of real space
 patterns of vortexes  in thermal equilibrium at a low temperature and b)
in nearby energy energy minima. The thermalized configurations shown in
 a) are obtained by performing Monte Carlo (MC) simulations at $T=0.2$
 on a system with $f=21/55$ and  $\lambda=1.5$. The equilibration
is extremely hard in this system so that we used a MC method which
 combines the Metropolis method, over-relaxation method and exchange MC
 method to endure equilibration \cite{MC-method}. 
The filled squares represent plaquettes with vortexes with charge $p=1$.
The configurations in b) are obtained by minimizing the energy by 
simple energy descent algorithm starting from the thermalized
 configurations shown in a). The energies of the energy minima of the configurations 1)-3) are $E=-5072.14311, -5072.38582, -5072.34445$ respectively.
} 
\label{fig-stripe-jja}
\end{figure}

Let us now turn to the {\it anisotropic} irrationally frustrated JJA with
$\lambda  > 1$.
Examples of the real space configurations of the vortexes
in equilibrium at a low temperature are shown in Fig.~\ref{fig-stripe-jja}. 
The most prominent feature is the stripe pattern
of the vortexes which are {\it regularly stacked} into $y$-direction (stronger
coupling) and {\it undulated} along the $x$-direction (weaker
coupling)
\cite{Yoshino-Nogawa-Kim-1,Yoshino-Nogawa-Kim-2,Yoshino-Nogawa-Kim-3}. The
formation of the stripes is reasonable because the repulsive
interactions between vortexes are anisotropic if $\lambda \neq 1$ as we
noted in sec \ref{subsubsec-vortex}.

There are two important observations.
First, the undulated stripe pattern is {\it frozen in time}, i.~e. the
ergordicity is broken. The pattern of the undulation cannot evolve dynamically
by usual relaxational dynamics once such a structure is established. 
This is simply because the stripes are {\it perfectly stacked} into the
$y$-direction in a belt. At a first sight, 
the stripe patterns may look similar to those found,
for example, in liquid crystals. But they are very different
because usual stripes fluctuate dynamically \cite{Chaikin-Lubensky}.

Second, there is a family of low lying states with
different patters of the transverse 
undulation as shown in Fig.~\ref{fig-stripe-jja}.
Apparently the ground state should have
no transverse undulation. Very interestingly the energies of the
different patterns of the undulation shown in Fig.~\ref{fig-stripe-jja}
are very close to each other suggesting a gap-less band of undulated states. 
Thus these undulated states are all relevant in the equilibrium ensemble.
This is manifested in the structure factor of the vortexes which
exhibit Bragg peaks into $q_{y}$ direction but a power law tail into
$q_{x}$ direction \cite{Yoshino-Nogawa-Kim-2,Yoshino-Nogawa-Kim-3}.

This is a very peculiar state of matter. Is this a glass?  ``No'', in
the sense that it has Bragg peaks which one would not expect for a
glass. ``Yes'', in the sense that there are many states with different
patterns of undulation, which is a self-induced disorder, and they are
separated by energy barriers. 

In a sense, the prediction by Halsey \cite{Halsey} - that superconducting glass (without quenched disorder) in the JJA with {\it irrational} $f$ - is realized. However we must keep in mind that here we are considering anisotropic  JJA with $\lambda > 1$ instead of the isotropic JJA $\lambda=1$  studied in most of the previous works. 

Now let us examine the low lying state more closer.
In the analysis of the ground state of the FK model, the hull function
\eq{eq-hull-FK} played a central role as we noted before. Since the JJA can be regarded as
a 2-dimensional version of the FK model, we are naturally led to look for similar one which may describe the low lying sates of the JJA in a
compact way. 

Because of the gauge invariance, let us focus
on the gauge-invariant phase differences across the Josephson couplings
$\psi_{ij}=\theta_{i}-\theta_{j}-A_{ij}$ defined \eq{eq-def-psi}
where $i$ and $j$ are nearest neighbours across a Josephson 
coupling which may be either along $x$ or $y$-axis. 
As we discuss later $\psi_{ij}$ is directly related to the Josephson
current $\sin(\psi_{ij})$, which is the analogue of {\it stress field}
in rheology.

\begin{figure}[h]
\begin{center}
\includegraphics[width=\textwidth]{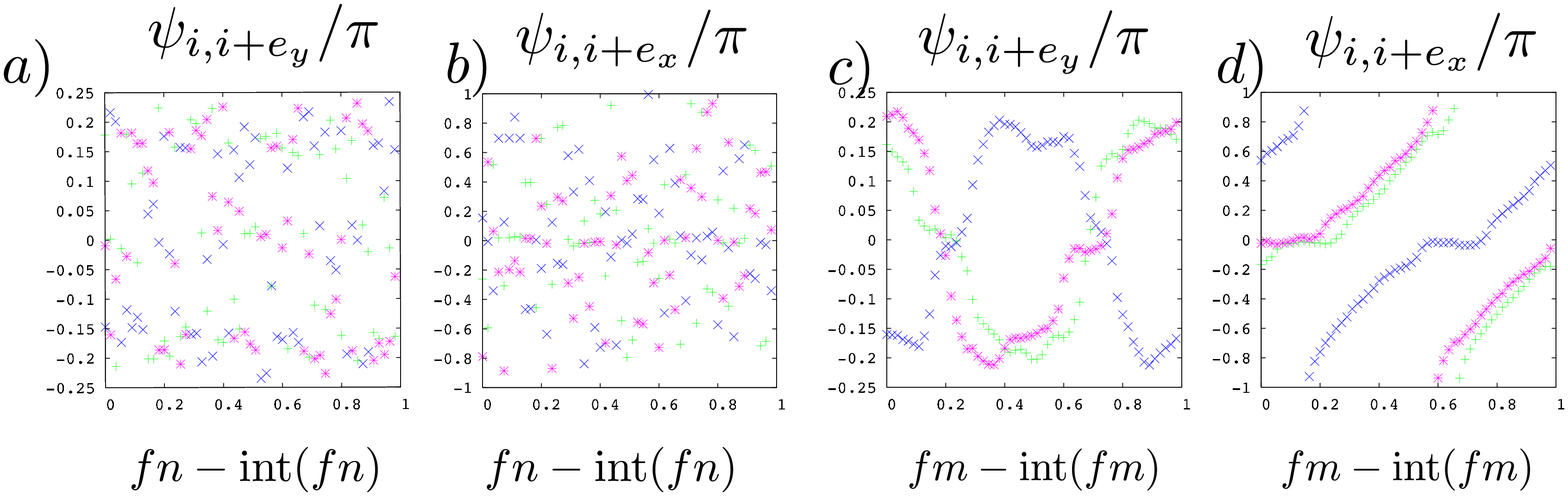}
\end{center}
\caption{(Color online) Spatial configuration of gauge invariant phases in an energy minimum. 
Phase differences $\psi_{i,i+e_{x}}$ and $\psi_{i,i+e_{y}}$ across
 Josephson junctions parallel to $x$ and $y$ axes at various sites
 $i=(n,m)$ are plotted against ``folded coordinates''.
Different symbols in the panels a) and b)
represent the phase differences at $n=1,2,\ldots,L$
along arbitrary chosen three different ``rows'' of the JJA.
In the panels c) and d), 
the phase differences at $m=1,2,\ldots,L$
along arbitrary chosen three ``columns'' are shown.
The system size is $L=55$. 
}  
\label{fig-hull-jja}
\end{figure}

In Fig.~\ref{fig-hull-jja} we display the phase differences $\psi_{ij}$
at various sites $i=(n,m)$ plotted against ``folded coordinates''
$[n]=fn-{\rm int}(fn)$ and $[m]=fm-{\rm int}(fm)$ which takes values limited in the range $0 < [n]  \leq 1$ and $0 < [m] \leq 1$.
The purpose of this plot is to elucidate the hull function 
analogously to the case of the FK model shown in Fig.~\ref{fig-hull-FK}.
Quite remarkably the plots in the panels c) and d) strongly suggest
there is indeed an analytic hull function of the folded coordinate along
the direction of {\it stronger coupling}. On the other hand, the panels
a) and b) suggest there are no such analytic hull functions along the
direction of {\it weaker coupling}. 

Recently we found it is possible to obtain the
hull functions analytically by performing a $1/\lambda$ expansion starting from 
$\lambda=\infty$ limit \cite{Yoshino-Nogawa-Kim-3}. 
It turned out that the transverse undulation is encoded in the
``phase differences'' between different columns which one can see
in the pane-ll c) and d).

The existence (absence) of analytic hull functions along stronger
(weaker) couplings immediately implies sliding (jamming) of the vortexes.
Starting from an energy minimum,  a family of different states with exactly
the same energy can be obtained through the operation $[m]\to
[m+\alpha]$ along the direction of stronger coupling with varying phase shift parameter $\alpha$. This amount to a unidirectional motion of the undulated vortex stripes into the
direction of stronger coupling without changing its pattern, i.~e. {\it sliding}.
In contrast, no such operation is possible along the direction of weaker
coupling, i.~e. {\it jamming}. In the next section we discuss how these
properties are reflected in physical observables associated with shear.

The above observation implies the symmetric system with $\lambda=1$,
on which most of the previous works have been dedicated, is actually
very special. As we discuss later, the critical point
corresponding to the Aubry's transition point is actually $\lambda_{c}=1$
in the JJA at zero temperature $T=0$.

\section{Response to shear - shear by external electric current}
\label{sec-shear}

Jamming is nothing but onset of rigidity which can be detected by response against shear. In general shear is induced into the system through the boundaries. In rheology one can consider to apply some constant external shear-stress $\sigma_{\rm ext}$ on boundaries of systems under study. Very interestingly this is equivalent to put external electric current $I_{\rm ext}$ into a Josephson junction array as we explain in sec \ref{subsec-dynamic-shear-response}. We will find that stress $\sigma$  and shear rate $\dot{\gamma}$  in rheology correspond precisely to current $I$ (or current density $J$) and voltage $V$ (or electric field $E$) in the transport problem of the driven JJA.

To study rheology (or transport) one can control either the
external stress $\sigma$ or shear rate $\dot{\gamma}$. Here we choose to control
the strain $\gamma$ so that we can discuss static and dynamic response to shear in the same set up. 

We put shear across the Josephson junction array along, say $y$-axis,
in the following manner. First we fix the configurations of the phase
variables $\theta_{i}$ on the bottom ($m=1$) and top ($m=L$) layers. 
Second we slightly change the boundary such that a uniform displacement
is imposed on the top layer ($m=L$) 
$\theta_{i} \to \theta_{i}+(L-1)\gamma$
while the bottom layer ($m=1$) is left in the
same fixed configuration. This amount to induce a gradient of phase
$d\theta/dy=\gamma$ along the $y$-axis. Clearly $\gamma$ corresponds to shear-strain of the usual sense. 
As the result some internal stress $\sigma$, which is super-current
running across the Josephson junctions (see below),  will be
induced in the system. To study rheology, we drive the top wall with a
constant speed so that the strain $\gamma$ increase with a constant 
shear-rate $\dot{\gamma}$. 
Let us also remark that shear $d \theta/d y=\gamma$ on the phases 
along the $y$-axis amount to motion of vortexes (dislocations) 
into the orthogonal direction, i.e. $x$-axis. This is equivalent to say that the vortexes are driven by the Lorentz force \cite{Tinkam}.

\subsection{Static response to shear - static rigidity}
\label{subsec-static-shear-response}

From static point of view, emergence of rigidity can
be best quantified by shear-modulus. 
The free-energy of the system $F(\gamma)$ 
can be formally expanded in power series of {\it infinitesimal}
shear strain $\gamma$ as
\beq
F(\gamma)=F(0)+N \langle \sigma \rangle \gamma + 
\frac{N}{2}\mu \gamma^{2} + \ldots
\eeq
where $\sigma$ and $\mu$ are the shear-stress and shear-modulus
respectively. $\langle \ldots \rangle$ stands for a thermal average.

Here $\gamma$ must be
infinitesimal. The free-energy density $F(\gamma)/N$, in the thermodynamic sense, must not
depend on the boundary condition (including the shape of the container) so that shear-modulus must be zero in
the thermodynamic sense  {\it even in solids}. Thus when the shear-modulus $\mu$
defined by the fluctuation formula \eq{eq-shear-modulus} emerges,
it means that the ordering of the $\gamma \to 0$ limit and the
thermodynamic limit $N \to \infty$ no more commute in sharp contrast to liquids.
In turn this means linear elasticity {\it must} fail in solids. Physically
this means that elasticity and {\it plasticity} must emerge
simultaneously in solids.\cite{YM2010}

It is useful to note that the change on the boundary condition can be formally ``absorbed'' into the bulk part of the system by replacing the original Hamlitonian \eq{eq-JJA-hamiltonian} by,
\beq
H(\gamma)= -\sum_{\vec{e}_{ij}=(1,0)} \cos(\psi_{ij})
 - \lambda \sum_{\vec{e}_{ij}=(0,1)} \cos(\psi_{ij}+\gamma).
\eeq
Based on this observation we find that the stress $\sigma$ can be expressed as,
\beq
N \sigma=\frac{\partial H(\gamma)}{\partial \gamma}
=\lambda \sum_{\vec{e}_{ij}=(0,1)} \sin(\psi_{ij}).
\label{eq-shear-stress}
\eeq
Similarly the shear-modulus $\mu$ can be expressed as,
\beq
\mu=b-\beta \left [ \langle \sigma^{2} \rangle - \langle \sigma \rangle^{2}
\right]
\label{eq-shear-modulus}
\eeq
where $b$ is {\it instantaneous} or adiabatic shear-modulus (``Born term'')
defined as,
\beq
N b=\frac{\partial^{2} H(\gamma)}{\partial \gamma^{2}}
=\lambda \sum_{\vec{e}_{ij}=(0,1)} \cos(\psi_{ij}).
\label{eq-born}
\eeq

Fluctuation formulae for the elastic modulus like \eq{eq-shear-modulus}
are well known in literature \cite{squire}. In the context of XY models
and super-conductors it is usually called as helicity modulus
\cite{Chaikin-Lubensky}. The crucial term is the 2nd term which
represents reduction of the shear-modulus due to thermal fluctuations of
the stress $\sigma$. 

In liquids, the two limits $\gamma \to 0$ and $N \to \infty$ should commute.
Then an identity $\mu=0$ must hold meaning exact cancellation must take place
between the Born term and the fluctuation term in \eq{eq-shear-modulus}.

In the previous section we found the anisotropic irrationally frustrated
JJA exhibits sliding/jamming in the low lying states such that the vortexes
can slide freely along the stronger coupling but jammed along the weaker
coupling. In turn this means that shear of the phases along
stronger/weaker coupling causes finite/zero changes of the energy
respectively. Consequently the shear-modulus $\mu$ must be finite/zero
along stronger/weaker coupling at zero temperature $T=0$. Indeed we observed this numerically \cite{Yoshino-Nogawa-Kim-1}.  

This is a quite intriguing situation - the anisotropic system $\lambda
\neq 1$ at zero temperature $T=0$ behaves either as solid or liquid
depending on the axes along which one imposes the shear. From numerical
observations it seems that the picture holds up to the symmetric point
$\lambda=1$ \cite{Yoshino-Nogawa-Kim-1} suggesting that the symmetric point
is actually the critical point $\lambda_{c}=1$ where shear-modulus
along a given axis becomes zero/finite.

\subsection{Dynamic response to shear - transport or rheology}
\label{subsec-dynamic-shear-response}

The shear-stress $\sigma$ defined in \eq{eq-shear-stress} is nothing but
super-current flowing along $y$-direction in the Josephson junction array. More precisely according to the  DC/AC Josephson relations \cite{Tinkam} the current $I_{ij}$  and voltage drop $V_{i}-V_{j}$ across the junction are given by,
\beq I_{ij}=\sin(\psi_{ij}) \qquad V_{i}-V_{j}=\frac{d\psi_{ij}}{dt}
\label{eq-dc-ac-Josephson-effect}
\eeq 
Here we are assuming some appropriate rescalings to define the
dimension-less quantities $I_{ij}$ and $V_{i}$.

At each site $i$ (super-conducting island) the current must be conserved.
By taking into account charging of the island and Ohmic energy dissipation we find,
\beq
C\frac{dV_{i}}{dt}+\sum_{j}\frac{V_{i}-V_{j}}{R}+\sum_{j}I_{ij}=I_{\rm ext}(\delta_{m_{i},L}-\delta_{m_{i},1})
\eeq
where the sums are took over nearest-neighbours. 
$C$ and $R$ are the capacitance of the islands and resistance of the
junctions respectively. 
$I_{\rm ext}$ is the strength of external current which is injected from
the top layer $m=L$ and extracted from the bottom layer $m=1$. Combining
with the Josephson relation \eq{eq-dc-ac-Josephson-effect} and the
definition of the gauge-invariant phase difference $\psi_{ij}$ given in
$\eq{eq-def-psi}$, one easily finds an effective equation of motion of the phases $\theta_{i}$, which is called RCSJ (Resistively and Capacitively Shunted Junction) model\cite{Tinkam}.
Apparently it can be cast into the form of Newton's equation of motion,
\beq
\frac{d\theta_{i}}{dt}=v_{i} \qquad
m \frac{d v_{i}}{dt}+\frac{\partial H}{\partial \theta_{i}}
+ \eta \sum_{j}(v_{i}-v_{j})=F_{\rm ext}(\delta_{m_{i},L}-\delta_{m_{i},1}).
\label{eq-motion}
\eeq
which can be considered as a toy model for rheology of layered systems under external shear applied on the top and bottom walls \cite{Yoshino}.

From the above observations, it is clear that transport properties in
JJA and rheology are quite analogous. Because of the shear, the velocity
field $d \theta_{i}/dt$ will acquire a slope along the $y$-axis which
can be identified with the shear rate $\dot{\gamma}$. From the AC
Josephson relation (the 2nd equation of \eq{eq-dc-ac-Josephson-effect}),
we find that it mounts to a constant electric field $E$ along the $y$ axis.

To summarize shear-stress $\sigma$ and shear-rate $\dot{\gamma}$ in
rheology correspond to electric current $I$ (or current density $J$) and voltage drop $V$ across
the system (or electric field $E$) in the transport problem of JJA. Thus
the so called ``flow curves'' in rheology corresponds to current-voltage $IV$ (or $JE$) characteristics in JJA.  In {\it tribology} we just need to 
consider only two layers $m=1,2$ as in the Matsukawa-Fukuyama's 2-chain
model \cite{Matsukawa-Fukuyama}. The yield stress $\sigma_{\rm c}$ is called as static frictional force.
These problems have been studied extensively in the
corresponding research communities but somehow the intimate analogy has not
been appreciated \cite{Yoshino}.

\subsection{Non-linear rheology and transport}

Let us discuss here some basic phenomenological aspects of the non-linear
rheology and the non-linear transport associated with 2nd order 
phase transition, including the jamming transition.
To be specific we will denote 
$\lambda-\lambda_{c}$ as the distance to the critical point which is
natural in the context of the anisotropic JJA at zero
temperature. However the readers can easily translate the discussion to
different situations by replacing $\lambda-\lambda_{c}$ by distance to
critical temperature $T-T_{c}$ or jamming density $\phi-\phi_{\rm J}$,
e.t.c. depending on the problems at hand.

Let us assume the following generic scaling form.
\beq
\sigma=|\lambda-\lambda_{c}|^{\beta}\tilde{\sigma}_{\pm}\left(\frac{\dot{\gamma}
}{|\lambda-\lambda_{c}|^{\Delta}}\right)
\label{eq-scaling}
\eeq
where $\beta$ and $\Delta$ are critical exponents
and the subscript $\pm$ stands for $\lambda > \lambda_{c}$  and $\lambda
< \lambda_{c}$ respectively. Physically we expect the following
behaviours: 1) Newtonian behaviour in the ``sliding phase''  ($\lambda <
\lambda_{c}$), 
2) Finite yield stress in the ``jammed phase'' ($\lambda > \lambda_{c}$) and
3) The explicit $\lambda$ dependence must disappear at the critical point ($J$-point) $\lambda=\lambda_{c}$. Based on these intuitions let us conjecture
the following asymptotic behaviours of the
scaling function $\tilde{\sigma}(y)$,
\beq
\tilde{\sigma}_{\pm}(y) = \left \{
\begin{array}{ll}

 \left \{
\begin{array}{cl}
y & \lambda < \lambda_{c}\\
\tilde{\sigma}_{-}(0) & \lambda > \lambda_{c} 
\end{array} \right. & y \ll 1  \\
 & \\
  c y^{\beta/\Delta}   & y \gg 1  
\end{array} \right.
\eeq
where $\tilde{\sigma}(0)$ and $c$ are some constants.
Consequently the scaling ansatz predicts the following asymptotic behaviours ($\dot{\gamma}\to 0$),
\beq
\lim_{\dot{\gamma} \to 0} \sigma = \left \{
\begin{array}{lll}
\eta(\lambda) \dot{\gamma}  &
\eta(\lambda)\propto (\lambda_{c}-\lambda)^{-(\Delta-\beta)} & \lambda < \lambda_{c}\\
c\; \dot{\gamma}^{\beta/\Delta} & & \lambda=\lambda_{c} \\
\sigma_{c}(\lambda)  & \sigma_{c}(\lambda) = \tilde{\sigma}_{-}(0)(\lambda-\lambda_{c})^{\beta} & \lambda > \lambda_{c}
\end{array} \right.
\eeq
Most importantly the power law fluid behavior $\sigma \propto \dot{\gamma}^{\beta/\Delta}$ at the critical point is predicted. Usually $\beta/\Delta < 1$ which is called {\it shear-thinning} behaviour.

For the transport problems in superconductors including JJA, one just
need to replace shear-stress $\sigma$ by the electric current density $J$, shear-rate $\dot{\gamma}$ by the electromagnetic field $E$. The Newtonian law corresponds to the Ohmic law 
$J=\sigma E$ with the linear conductivity $\sigma$\footnote{It should not be confused with stress $\sigma$} and the yield stress $\sigma_{c}$ corresponds to critical current $J_{c}$.

Recently the non-liner rheology of granular systems is found to obey
this type of scaling around the J-point
\cite{Otsuki-Sasa,Hatano-Otsuki-Sasa,Olsson-Teitel,Hatano,Otsuki-Hayakawa}. In
granular systems Bagnold's scaling must replace the Newtonian law in the unjammed phase. At least formally, the above argument can be easily modified to account for it.

The above scaling ansatz is quite reminiscent of the scaling property of
magnetization of ferromagnetic models around the critical temperature $T_{c}$. 
On purpose we actually used the same standard notations for the critical
exponents, i.~e. $\beta$ and $\Delta$, in the latter problem. Namely by
replacing the stress $\sigma$ by magnetization $m$ and strain rate $\dot{\gamma}$ by magnetic field $h$, one recovers $m \propto |T-T_{c}|^{\beta}\tilde{m}_{\pm} (h/|T-T_{c}|^{\Delta})$. One can easily find precise correspondences between  1) the Newtonian (Ohmic) law v.s. paramagnetic behaviour $m=\chi h$ with the linear-susceptibility $\chi$ diverging at $T_{c}$ 2) power law rheology $\sigma \propto \dot{\gamma}^{\beta/\Delta}$ v.s.
$m \propto h^{\delta}$ with $\delta=\beta/\Delta$ at the critical points and 3) yield stress (critical current)
$\sigma_{c} \propto (\lambda-\lambda_{c})^{\beta}$ v.s. the spontaneous magnetization $m_{s} \propto (T_{c}-T)^{\beta}$.

This type of scaling has been advocated first in the context of non-linear current-voltage characteristics of superconductors by Wolf, Gubser and Imry \cite{WGI}.
They studied non-linear current-voltage characteristics of superconducting film at
the superconducting phase transition, which is a Kosterlitz-Thouless
type 2nd order phase transition\cite{KT}.  They pointed out
the analogy with the scaling of the magnetization of ferromagnets. Such 
dynamical scaling ansatz has been extensively used in the studies of
transport properties in high-$T_{\rm c}$ superconductors, especially in the
context of the {\it vortex-glasses}  with quenched pinning 
centers \cite{FFH}.

More recently Otsuki and Sasa \cite{Otsuki-Sasa} has realized the same
type of critical behavior in the context of the non-linear rheology of
molecular glasses. Quite remarkably they were able to find a mean-field theory
which predicts that the flow curves of the non-linear rheology are
formally identical to the equation of state of the Landau-Ginzburg
theory of ferromagnets under external magnetic field suggesting in
particular $\beta/\Delta=1/3$.
 

\begin{figure}[t]
\begin{center}
\includegraphics[width=0.75\textwidth]{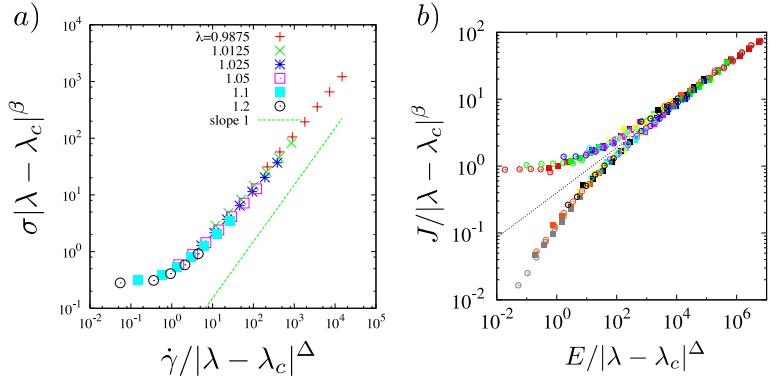}
\end{center}
\caption{(Color online)
Flow curves of the FK model and JJA. Both models are simulated by the RCSJ method at zero temperature $T=0$.
The panel a) displays the master flow curve of the FK model using 
$\lambda_{c}=0.9716$..and $\beta=\Delta=3$. The panel b) displays the master flow curve of the irrationally frustrated anisotropic JJA using $\lambda_{c}=1$, $\beta=1.19$ and $\Delta=3.5$ so that $\beta/\Delta=0.34$.
} 
\label{fig-flowcurves}
\end{figure}

Let us now discuss the dynamical scaling properties of the FK model and the JJA under shear.
We performed the RCSJ simulation on both models.
The master flow curve of the FK model is displayed in the panel a) of Fig.~\ref{fig-flowcurves} which follow well the expected dynamical scaling behaviour around the Aubry's transition point $\lambda_{c}$. 
A previous work \cite{Peyrard-Aubry} found $\sigma_{c} \propto (\lambda-\lambda_{c})^{\tilde{\psi}}$ with $\lambda_{c}=0.9716..$ and $2.85 < \tilde{\psi} < 3.06$.  In addition we found the system remains Newtonian for the entire sliding phase $\lambda \leq \lambda_{c}$ including the critical point so we assumed $\beta=\Delta$ in the scaling plot.

For the irrationally frustrated anisotropic JJA, 
we pointed out in sec \ref{subsec-static-shear-response} that the shear-modulus $\mu$ along stronger/weaker
coupling is finite/zero at zero temperature $T=0$ and that the symmetric point $\lambda_{c}=1$ is the critical point where the shear-modulus $\mu$ along a given axes changes from finite/zero to zero/finite. 
Then it is quite natural to expect that the current-voltage curve of the system with respect to
injection of the electric current along a given direction exhibit dynamical scaling feature at around $\lambda_{c}=1$. This is indeed observed by a numerical simulation of the RCSJ dynamics \cite{Yoshino-Nogawa-Kim-1}.
The current-voltage curves collapse onto a master curves as shown in the panel b) of Fig.~\ref{fig-flowcurves}.

\begin{figure}[t]
\begin{center}
\includegraphics[width=0.5\textwidth]{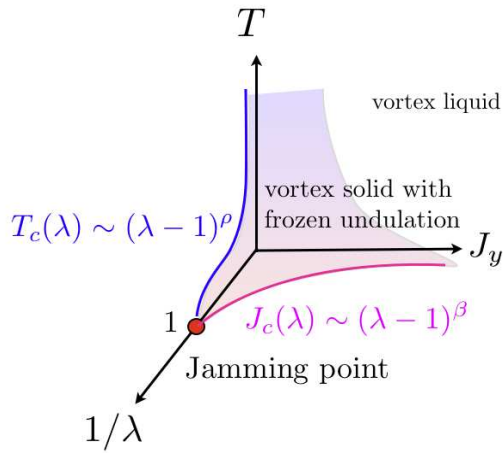} 
\end{center}
\caption{(Color onlie) ''Jamming'' phase diagram of the JJA  
}  
\label{fig-phasediagram}
\end{figure}

\subsection{Jamming phase diagram}

In Fig.~\ref{fig-phasediagram} we show a schematic ``jamming phase diagram'' of the irrationally frustrated anisotropic Josephson junction array. So far we discussed only the properties of the system at zero temperature $T=0$ and $\lambda \geq 1$. Under the electric current $J_{y}$ injected along the $y$-axis, the system remains jammed as long as $J_{y}$ is smaller than the critical current $J_{\rm c} \sim  (\lambda-1)^{\beta}$. The configuration of the jammed solid phase is characterized by the frozen pattern of the undulated vortex stripes. Under strong enough current $J_{y} > J_{\rm c}$, the system starts to move exhibiting the shear-thinning behaviour. On the other hand the same solid can slide freely with respect to electric current $J_{x}$ injected along the $x$-axis for the entire $\lambda > 1$ region. For the region $\lambda < $, we just need to interchange the $x$ and $y$-axes in the above discussion.

Previous studies on the irrationally frustrated JJA are almost exclusively concerned with the symmetric point $\lambda=1$. Recent intensive numerical studies at finite temperatures suggest $T_{c}(\lambda=1)=0$  \cite{Park-Choi-Kim-Jeon-Chung,Granato} without finite temperature glass transition anticipated in the early works \cite{Halsey,JJA-relaxation}. On the other hand our recent studies on the static properties at low temperatures \cite{Yoshino-Nogawa-Kim-2,Yoshino-Nogawa-Kim-3} strongly suggest $T_{c}(\lambda) > 1$ at least sufficiently away from the symmetric point $\lambda=1$ and that  $T_{c}(\lambda)$ rapidly decreases approaching the symmetric point $\lambda=1$. These point toward the possibility of the jamming phase diagram depicted in  Fig~\ref{fig-phasediagram}.
\footnote{We only show the $\lambda >1$ part. Note that $T_{c}(1/\lambda)=T_{c}(\lambda)/\lambda$ holds
due to the obvious symmetry between $x$ and $y$ axis. }
Quite interestingly it is very similar to the jamming phase diagram proposed by the Chicago group \cite{Liu-Nagel,Nagel-group}. Most notably the point $(\lambda=1,T=0)$ look quite similar to the Jamming point which deserves further studies.

\section{Conclusions}
\label{sec-conclusions}

To conclude we discussed static and transport or rheological properties of the irrationally frustrated anisotropic Josephson junction array (JJA) which exhibits vortex stripes with self-generated randomness at low temperatures. We emphasized in particular the intimate connection between the friction models and the irrationally frustrated JJA which provides valuable insights into the problems.

\section*{Acknowledgements} We thank Takahiro Hatano, Hikaru Kawamura, Jorge Kurchan
and Hiroshi Matsukawa for useful discussions.
We thank the Supercomputer Center, ISSP, University of Tokyo for the use of the facilities.
This work is supported by Grant-in-Aid for Scientific Research
on Priority Areas "Novel States of Matter Induced by Frustration"
(1905200*) and Grant-in-Aid for Scientific Research (C) (21540386).

%

\end{document}